\documentclass[letterpaper,10pt]{article} 

\usepackage{osameet3} 

\usepackage{amsmath,amssymb}
\usepackage[colorlinks=true,bookmarks=false,citecolor=blue,urlcolor=blue]{hyperref} 
\usepackage{caption}
\usepackage{subcaption}
\newcommand*{\affmark}[1][*]{\textsuperscript{#1}}

\begin{document}

\title{A Novel low-latency DBA for Virtualised PON implemented through P4 In-Network Processing}

\vspace{-4mm}

\author{Diego Rossi Mafioletti\affmark[1]\affmark[3], Frank Slyne\affmark[1], Robin Giller\affmark[2], Michael O'Hanlon\affmark[2], \\
Brendan Ryan\affmark[2] and Marco Ruffini\affmark[1]}
\address{\affmark[1]CONNECT Centre, Trinity College Dublin, \affmark[2]Intel Corporation, Ireland, \affmark[3]Federal Institute of Esp{\'i}rito Santo}
\email{\{rossimad,fslyne,marco.ruffini\}@tcd.ie, \{robin.giller, michael.a.ohanlon, brendan.ryan\}@intel.com}

\copyrightyear{2021}

\vspace{-3mm}
\begin{abstract}
We present a novel dual-DBA allocation, with a fast P4-enabled scheduler to provide low latency upstream grant allocations. We show latency reduction of 37\% and 43\%, respectively, compared to standard and virtual PONs.
\end{abstract}
\vspace{-0.8mm}
\section{Introduction}
Passive Optical Networks (PONs) are considered a cost-effective architecture for ubiquitous broadband delivery, due to their ability to share cost and capacity across end points. For this reason, they are being increasingly considered as a possible solution to connect small cells in 5G functional split architecture (i.e., supporting ORAN 7.2 Remote Unit(RU)-Distributed Unit(DU) split as well as higher level splits).
One of the main drawbacks of the PON point-to-multipoint topology is upstream latency, which is higher compared to simpler point-to-point solution, as the scheduling mechanism requires exchange of reports and bandwidth map calculations that introduce  additional few hundreds microsecond delay. 
Approaches such as Cooperative DBA \cite{coDBA}, recently standardised as Cooperative Transport Interface (CTI) \cite{CTI}, have addressed this latency issue by providing coordination between mobile scheduling at the DU and OLT scheduling. However, CTI works because in Cloud-RAN the low latency issue is generated by a protocol mismatch between PON and RAN rather than by application-level requirements. For this reason, the CTI is able to fix the issue by sharing the advanced scheduling information from the DU with the OLT. This exchange of information is further facilitated by PON virtualisation mechanisms \cite{Ruffini2020}, which simplifies the integration between wireless and optical technologies.

In this paper we address a more challenging issue, where the low latency requirement comes from the application. This means that neither an OLT nor a DU (in case the application runs over a mobile network) can know in advance when an upstream transmission request will arrive at the ONU.
Currently, the only known working solution is to assign a fixed upstream allocation to a given ONU, which is thus allowed to transmit a given number of bytes, potentially every frame, without requesting a grant allocation. While this mechanism does provide the lowest latency, it is highly inefficient as capacity is statically assigned to the ONU. For this reason it is also not scalable. Other mechanisms to reduce latency are based on prediction of traffic arrival \cite{Mondal}. These however need to be tuned application by application and can only work for specific applications that present well defined packet arrival patterns that can be predicted several hundreds of $\mu s$ in advance (their arrival time also needs to be estimated with sub-microsecond precision).
A well performing low latency PON algorithm was presented in \cite{NTT}, however this only applies to Ethernet PON standard, while in this work we focus on the ITU-T standard.

In this paper we propose a novel mechanism, that splits the upstream DBA scheduling in two parts. One that operates according to standard DBA procedures, where the bandwidth map is calculated after all required grant requests (Dynamic Bandwidth Report units - DBRus) are received for a given cycle. Here, the only difference is that our mechanism requires that part of this bandwidth map is initially left unallocated.
A second mechanism, which we call Fast Intercept, operates independently a faster grant allocation that updates the standard bandwidth map (BWMAP) before it's sent to all ONUs, to provide immediate grants to newly arrived DBRu messages that are associated with low-latency service (i.e., without waiting for an entire DBA cycle). 

One of the key features of our implementation, which is optimised for virtual PON architectures, is that the standard DBA runs as a Virtual Network Function (VNF) on a general purpose processor (i.e., in the server running the virtual OLT and other network VNFs), while the Fast Intercept mechanism runs in the network card, operating the low-latency grant processing. In our implementation, this is carried out on a programmable P4 \cite{Bosshart2014} pipeline.


\section{Low-latency DBA description and implementation}
\begin{figure}[htb]
    \centering
    \begin{minipage}{0.5\linewidth}
        \begin{subfigure}[t]{0.4\linewidth}
            \centering
            \includegraphics[scale=0.28]{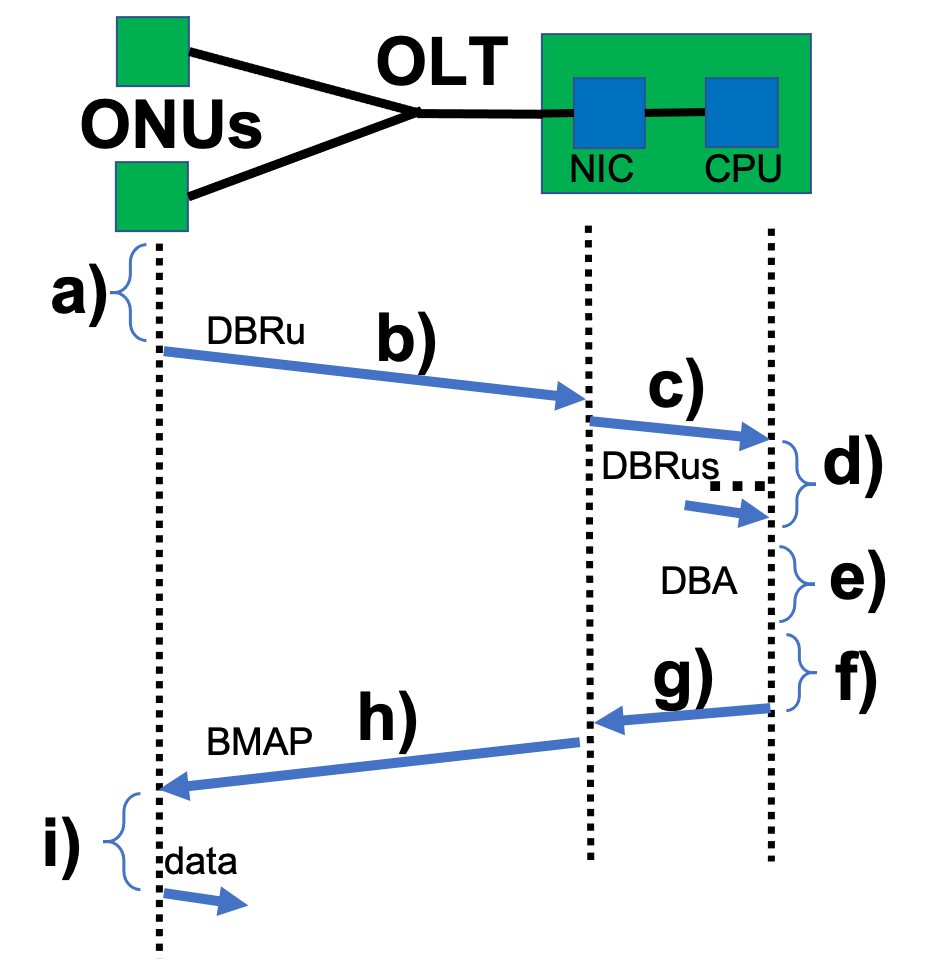}
            \caption{Standard DBA process.}
            \label{fig:standardDBA}
        \end{subfigure}
        \hfill
        \centering
        \begin{subfigure}[t]{0.4\linewidth}
            \centering
            \includegraphics[scale=0.28]{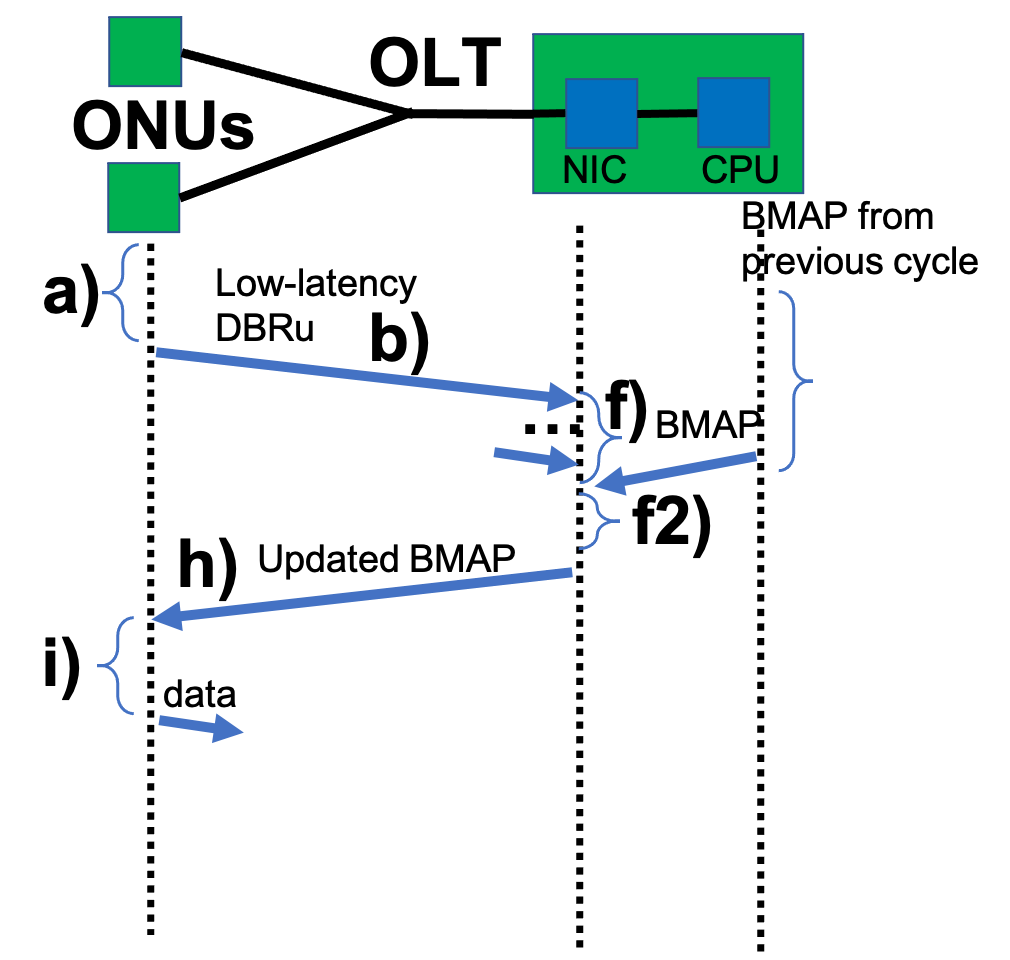}
            \caption{Proposed dual-DBA process with P4 in-NIC computing.}
            \label{fig:proposedDBA}
        \end{subfigure}
        \vspace{-3mm}
        \caption{Summary of steps involved in DBA process.}
    \end{minipage}
    \hfill
    \begin{minipage}{0.35\linewidth}
        \begin{subfigure}[t]{0.5\textwidth}
            \centering
            \includegraphics[scale=0.5]{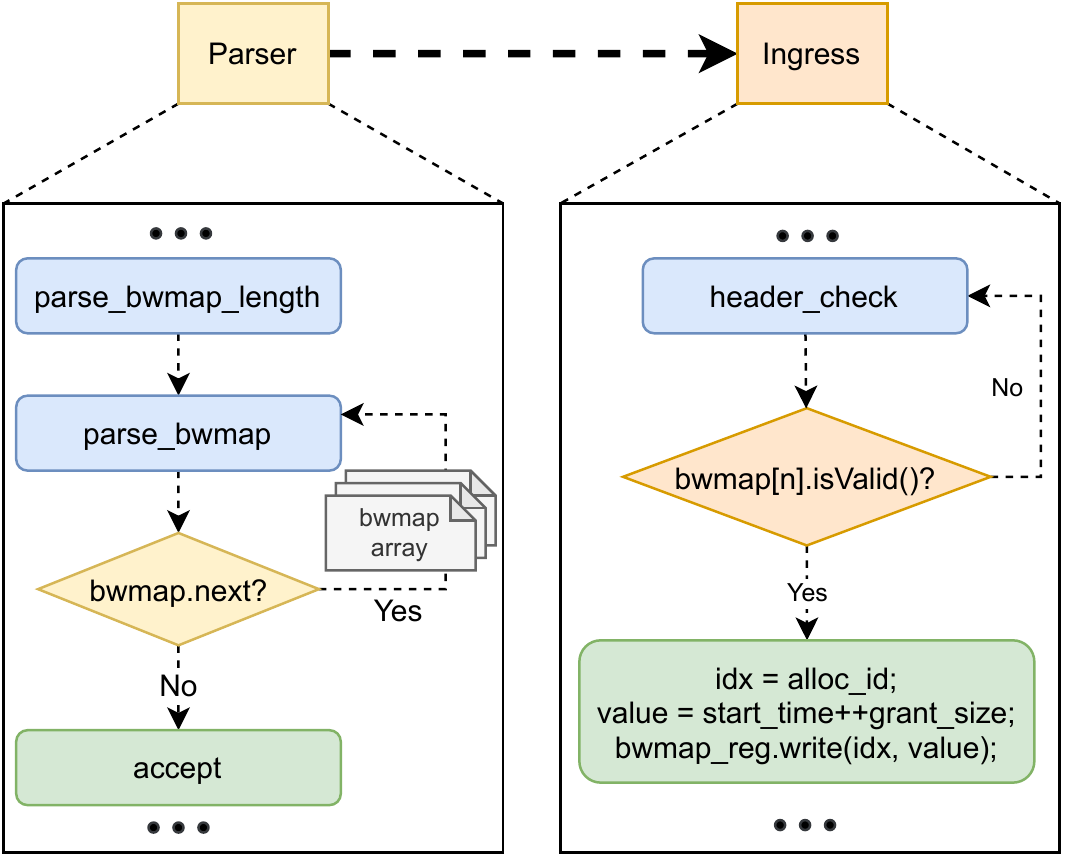}
        \end{subfigure}
        \caption{P4 Parser and Ingress pipelines}
        \label{fig:p4Pipelines}
    \end{minipage}
    \vspace{-5mm}
\end{figure}

Figure \ref{fig:standardDBA} reports the different steps involved in a DBA process, together with typical latency times, from the moment a packet arrives at the ONU queue, until the moment that ONU is allowed to transmit the packet. Some of the latency times are typical of DBA implementation, while others are experimentally measured in our setup (and further discussed in section 3 below). a) the ONU needs to wait for the opportunity to piggy back the DBRu to an upstream message (between 0 and 125 $ \mu s$, for an average of 62.5 $ \mu s$); b) the DBRu propagates trough fibre (assume 50 $ \mu s$ for a 10 km distance); c) the information travels between the physical card and the virtual process (this only occurs for virtual PON implementations, about 22 $ \mu s$ from our experimental data, .i.e. half the round trip time of 41.96 $\mu s$); d) the DBA process waits for a given time window to receive DBRus from multiple ONUs (between 0 and 125 $ \mu s$, for an average of 62.5 $ \mu s$); e) the OLT runs the DBA algorithm to calculate the Bandwidth Map (assume DBA calculation time of 77 $ \mu s$, according to results in figure \ref{fig:hardwareVSsoftware} - difference between second and third bars in the plot); f) the Bandwidth Map is included at the beginning of the next downstream frame (between 0 and 125 $ \mu s$, for an average of 62.5 $ \mu s$); g) the Bandwidth Map travels between the virtual functional and the physical card (same consideration as c); h) the Bandwidth Map propagates trough fibre (same considerations as b); i) the ONU can transmit the data at its allocated time (considering we can schedule low latency allocations at the beginning of a frame, we assume between 0 and 20 $ \mu s$, for an average of 10 $ \mu s$). 

Our proposed approach is illustrated in Figure \ref{fig:proposedDBA}. The main difference here is that the grant calculations for the Fast Intercept mechanism occur in parallel in the P4 NIC, while waiting for a BWMAP to arrive from the CPU VNF. We assume that the T-CONT ID is used to determine whether an allocation requires low-latency support. As soon as it arrives, the Fast Intercept mechanism modifies the BWMAP to include the latest arrival low-latency grant requests.
Thus, with respect to the stages above (we adopt the same step labels for ease of comparison), we have: a) The ONU needs to wait for the opportunity to piggy back the DBRu to an upstream message; b) the DBRu propagates through fibre; f) the next Bandwidth Map arrives at the NIC (between 0 and 125 $ \mu s$, for an average of 62.5 $ \mu s$), in parallel the NIC calculates the BWMAP update for the low latency grants (7.55 $\mu s$ according to our results in section 3); f2) the NIC updates the BWMAP including the low latency grant allocations (2.5 $\mu s$ according to our results in section 3); h) the BWMAP propagates through fibre; i) the ONU can transmit the data at its allocated time.

\vspace{-2mm}

Considering the calculations provided above, the minimum average time for low latency allocation through a classical DBA mechanism is of 374.5 $ \mu s$, which is increased to 418.5 $ \mu s$ for a virtual implementation (i.e. considering of the additional steps c) and g) above).
On the other hand, the proposed Fast Allocation mechanism, under similar conditions, can reduce this value to 237 $ \mu s$. This values represent a reduction of 37\% and 43\%, respectively, compared to a classical PON and virtual PON implementation.


Before going into further implementation details, one consideration we want to make is whether allocating part of the bandwidth map for low latency applications could be considered wasteful, in case there are not enough applications requiring it on any given frame. In our implementation we easily solve this issue: 
as all DBRu request always pass through the P4 NIC, the Fast Intercept mechanism can fill in spare allocations using requests from other lower priority grant requests (i.e., that do not have low latency constraints), to avoid wasting capacity.
The opposite issue could also occur, that the unallocated bandwidth map is not large enough to accommodate all low latency requests. In this case we implement a policy where the Fast Intercept mechanism could preempt additional allocations in the BWMAP that are currently associated to best effort services. 

\textbf{\textit{P4 implementation details}}
We developed a P4 embedded network function (eNF) on a Netronome SmartNIC that processes the BWMAPs and DBRus data structures \cite{Mafioletti2020}. Figure \ref{fig:p4Pipelines} shows the P4 Parser and Ingress pipelines on the SmartNIC. 
For each upstream burst arriving to the SmartNIC, the eNF checks the T-CONTs and stores their content (DBRus) in a data structure for subsequent checking in the downstream direction. Thus, on the downstream direction, the eNF analyses the data stored previously, runs a simplified fast allocation DBA algorithm on the network hardware and updates the upcoming BWMAP accordingly.

The algorithm modifies the start\_time and grant\_size in the BWMAP. As mentioned above, the system starts with part of the BWMAP reserved for low latency traffic. The process then follows the following 4 steps: \\
\vspace{-6mm}
\begin{enumerate}
\item The eNF identifies the  DBRus in transit into a P4 register, which is a fast memory on the network hardware. 
\vspace{-6mm}
\item If the DBRu \textbf{does not} include a low latency grant request, the DBRu packets continues to its destination (the DBA in the CPU), without any modification.
\vspace{-3mm}
\item If the DBRu \textbf{does} include a low latency grant request, the eNF prepares the allocation that will be used to modify the incoming BWMAP, applying any required modification to the grant\_size fields.
\vspace{-3mm}
\item When the next BWMAP arrives to the NIC from the CPU, the P4 process modifies this accordingly to include the low latency allocations, before forwarding it to the ONUs. 
\vspace{-2mm}
\end{enumerate}
\section{Experimental Results}





\begin{figure}[htp]
    \centering
    
    \begin{subfigure}[b]{0.45\textwidth}
        \centering
        \includegraphics[scale=0.45]{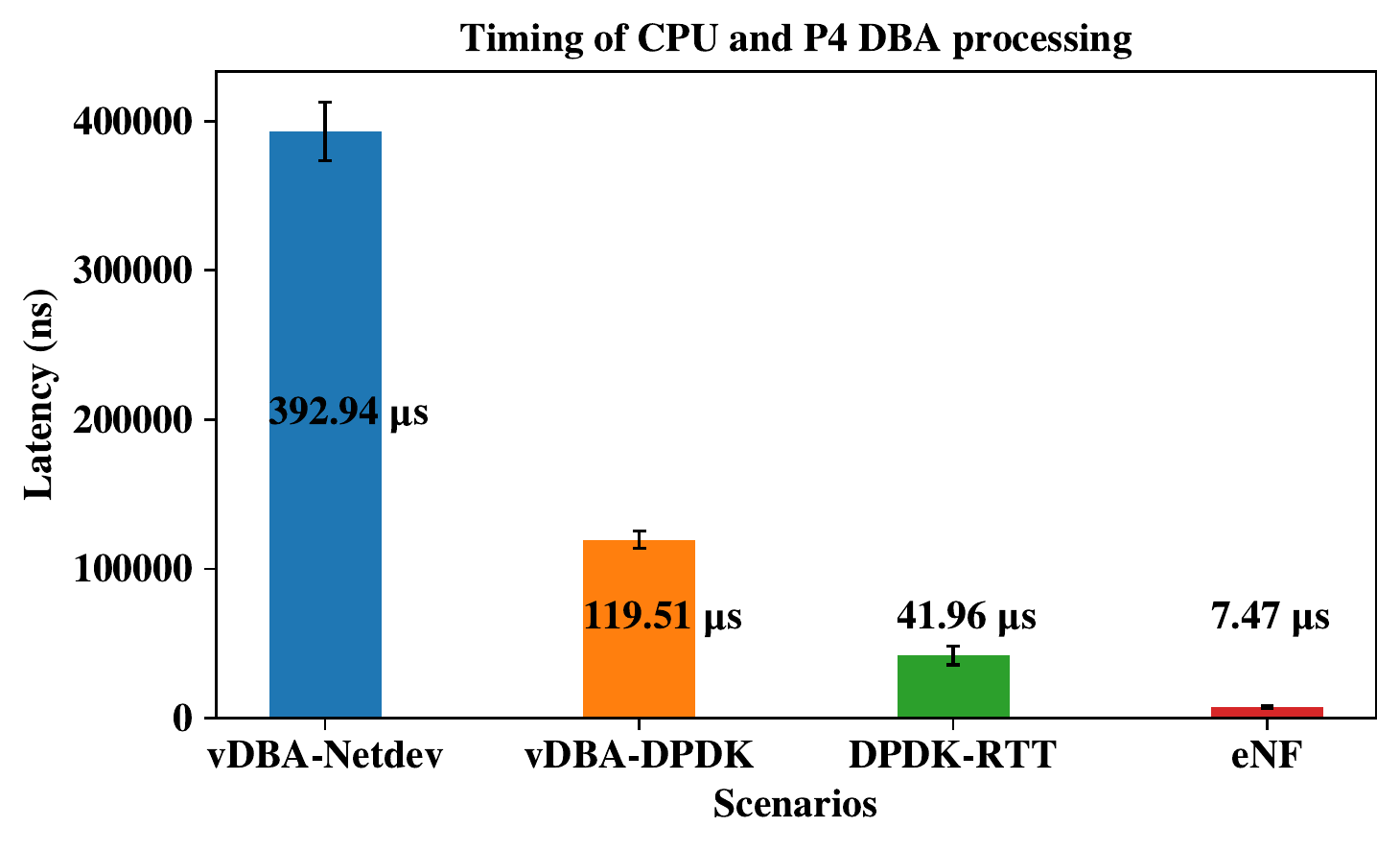}
        \vspace{-3mm}
        \caption{Comparison between hardware/software}
        \label{fig:hardwareVSsoftware}
    \end{subfigure}
    \begin{subfigure}[b]{0.45\textwidth}
        \centering
        \includegraphics[scale=0.45]{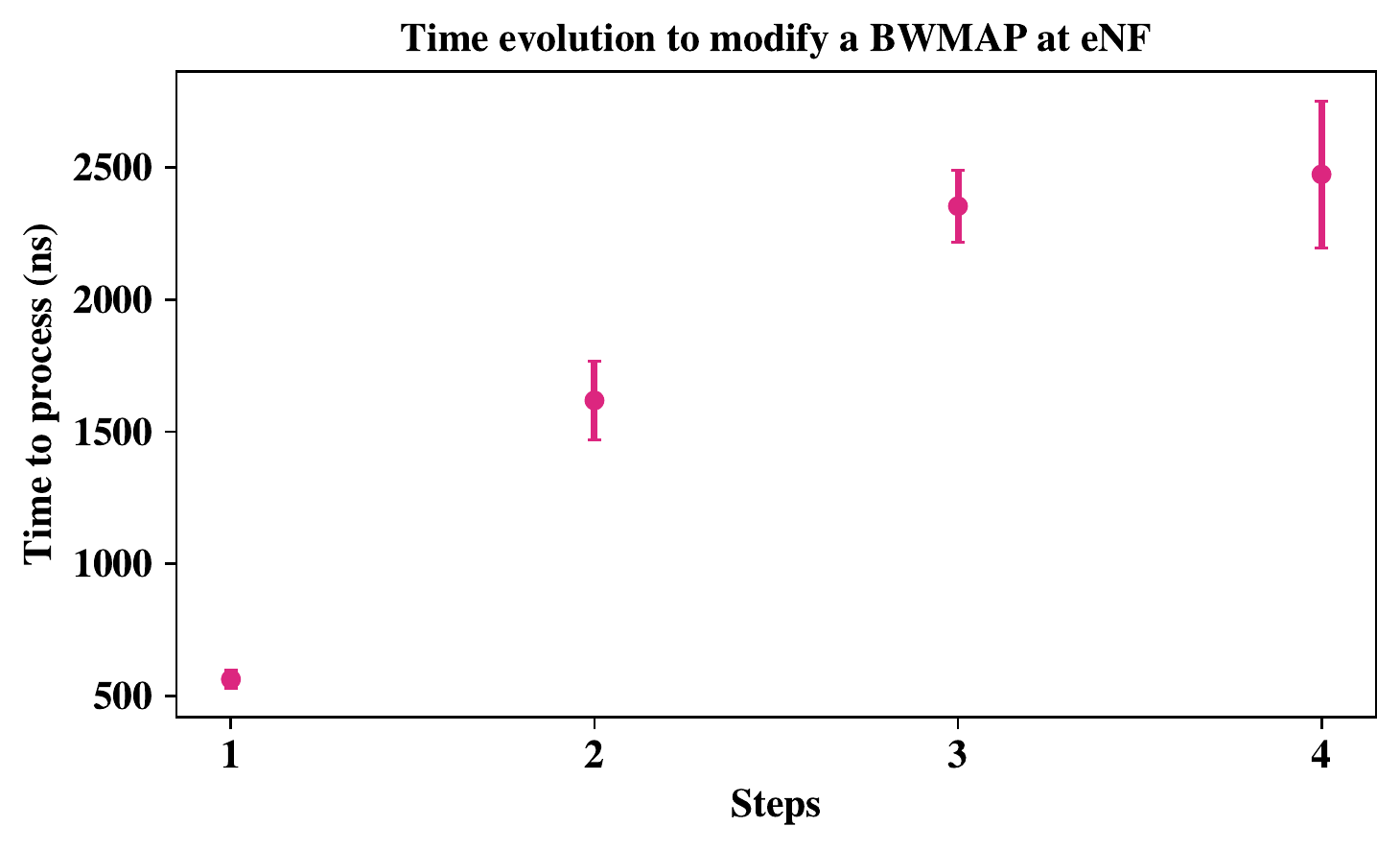}
        \vspace{-3mm}
        \caption{Dissecting the algorithm: times of each step.}
        \label{fig:dissection}
    \end{subfigure}
    \vspace{-3mm}
    \caption{Timing analysis of the CPU and P4 DBA processing}
    \label{fig:timeToProcess}
    \vspace{-10mm}
\end{figure}

Figure \ref{fig:hardwareVSsoftware} reports both DBA computation time and transmission time between SmartNIC and CPU. These affects the steps c), e) and g) shown in Fig. 1. The first bar in Fig. \ref{fig:hardwareVSsoftware} reports the time required to send DBRus from NIC to CPU, for the DPU to calculate the BWMAP and for this to be sent from the CPU to the NIC. This is a baseline scenario, where no specific optimisation is carried out, using Linux Netdev, and show the longest time of about 393 $\mu s$. 
The second bar represents the same process, but when implemented through our optimised DPDK solution for virtual PON. This bypasses the Linux network stack, and runs in user space in poll mode, and it reduces the timing to 119.51 $\mu s$. The third bar shows the time required for a round trip time between NIC and CPU when using DPDK (41.96 $ \mu s$). From the difference we can infer the DBA processing time in the DPDK implementation of 77.55 $\mu s$.
The forth bar finally, shows the time required by the NIC to operate the Fast Intercept mechanism, inclusive of grant calculation and update of the BWMAP. As this does not require any data transfer between NIC and CPU and allows fast P4 processing for the fast DBA, it can be computed is only 7.47 $\mu s$. It should also be noted that since the PON is a synchronous TDM technology, the eNF knows exactly when the BWMAP will be received form the CPU. Thus it can initiate the fast DBA calculation 8 $\mu s$ before it receives the BWMAP. In this case the only additional time will be that required to modify the BWMAP.
This is reported in Fig. \ref{fig:dissection}, which shows the break down of the eNF computation time (i.e., the measured timings of the four steps described at the end of section 2). We can see that the time required to modify the BWMAP is of the order of 2.5 $\mu s$, which is thus the only additional time by which the BWMAP is delayed after arriving at the NIC.

Summarising the experimental results in Fig. \ref{fig:hardwareVSsoftware} and \ref{fig:dissection}, re-iterating the overall DBA calculation times reported in section 2 and Fig. \ref{fig:standardDBA} and \ref{fig:proposedDBA}, we have calculated that a standard DBA mechanism (assuming the DBA is calculated for each service interval), would have an average latency of 374.5 $ \mu s$ and  418.5 $ \mu s$, respectively for an OEM and virtual implementation of a PON DBA. On the other hand our proposed mechanism, split into a CPU and SmartNIC implementation can provide an average latency of 237.5 $ \mu s$. This is a significant reduction of 37\% and 43\%, for upstream PON latency, which can enhance the PON support for low latency applications.

\vspace{-2mm}

\section*{Acknowledgements}
\vspace{-0.5mm}
\small
Financial support from Science Foundation Ireland grants 14/IA/2527 and 13/RC/2077 is gratefully acknowledged.


\end{document}